\documentclass[aps,prc,reprint,superscriptaddress,floatfix]{revtex4-1}

\usepackage{CJKutf8}
\usepackage{graphicx}
\usepackage{amsmath}
\usepackage{amsfonts}
\usepackage{xcolor}
\usepackage[colorlinks,linkcolor=blue,urlcolor=blue,citecolor=blue]{hyperref}

\begin{document}
\begin{CJK}{UTF8}{gbsn}



\title{Machine-learning-based identification for initial clustering structure in relativistic heavy-ion collisions}


\author{Junjie He(何俊杰)}
\affiliation{Shanghai Institute of Applied Physics, Chinese Academy of Sciences, Shanghai 201800, China}
\affiliation{University of Chinese Academy of Sciences, Beijing 100049, China}

\author{Wan-Bing He(何万兵)}
\email{hewanbing@fudan.edu.cn}
\affiliation{Key Laboratory of Nuclear Physics and Ion-beam Application (MOE), Institute of Modern Physics, Fudan University, Shanghai 200433, China}

\author{Yu-Gang Ma(马余刚)}
\email{mayugang@fudan.edu.cn}
\affiliation{Key Laboratory of Nuclear Physics and Ion-beam Application (MOE), Institute of Modern Physics, Fudan University, Shanghai 200433, China}

\author{Song Zhang(张松)}
\affiliation{Key Laboratory of Nuclear Physics and Ion-beam Application (MOE), Institute of Modern Physics, Fudan University, Shanghai 200433, China}

\date{\today}

\begin{abstract}
$\alpha$-clustering structure is a significant topic in light nuclei.
A Bayesian convolutional neural network (BCNN) is applied to classify initial non-clustered and clustered configurations, namely Woods-Saxon distribution and three-$\alpha$ triangular (four-$\alpha$ tetrahedral) structure for $^{12}$C ($^{16}$O), from heavy-ion collision events generated within a multi-phase transport (AMPT) model.
Azimuthal angle and transverse momentum distributions of charged pions are taken as inputs to train the classifier.
On multiple-event basis, the overall classification accuracy can reach $95\%$ for $^{12}$C/$^{16}$O + $^{197}$Au events at $\sqrt{S_{NN}} =$ 200 GeV.
With proper constructions of samples, the predicted deviations on mixed samples with different proportions of both configurations could be within $5\%$.
In addition, setting a simple confidence threshold can further improve the predictions on the mixed dataset.
Our results indicate promising and extensive possibilities of application of machine-learning-based techniques to real data and some other problems in physics of heavy-ion collisions.
\end{abstract}


\maketitle
\end{CJK}


\section{\label{SecIntro}Introduction}
Clustering is an extensively existing phenomenon in nuclei.
Especially in light nuclei, the mean field effect is not strong enough to break cluster structure, leading to possible observations of clustering behaviors in the excited states or even in the ground state.
$\alpha$ cluster plays the most critical role in light nuclei clustering due to its extreme tightness and the strong repulsive $\alpha$-$\alpha$ interaction~\cite{shimodaya1962interaction,endo1964phenomenological}.
$^{12}$C is of particular importance because its formation from $^{4}$He is a bottleneck in the process of stellar nucleosynthesis~\cite{Jin,LiWJ,Tang}.
Although the Hoyle state was predicted and confirmed in the 1950s~\cite{hoyle1954nuclear,dunbar19537,cook1957b}, its precise properties are still being investigated.
An obtuse triangular configuration of $\alpha$ clusters is predicted from an $ab~initio$ lattice simulation using effective field theory~\cite{epelbaum2011ab,epelbaum2012structure}.
As for the ground state, antisymmetrized molecular dynamics (AMD) shows the $p_{3/2}$ sub-shell closure configuration with mixing of the 3$\alpha$-core component~\cite{kanada2007structure}.
$ab~initio$ lattice calculations also support a compact triangular configuration of $\alpha$ clusters~\cite{epelbaum2012structure}.
Moreover, the later-discovered high-spin $J^{\pi}=5^-$ state fits well to the ground-state rotational band predicted by the algebraic cluster model (ACM).
In this model, the three-$\alpha$ configuration is arranged on an equilateral triangle, indicating a $\mathcal{D}_{3h}$ symmetry~\cite{marin2014evidence}.

In the viewpoint of nuclear reaction, some observables have been proposed as probes of $\alpha$ clustering.
For example, the giant dipole resonance spectrum~\cite{he2014giant,he2016dipole,Huang2021}, collective flows~\cite{
guo2019influence,NST_Shi}, and the nucleon emission from photon-induced reactions in quasi-deuteron region~\cite{huang2017photonuclear,Huang2020} were demonstrated as potential probes for $\alpha$-clustering nuclei in low-intermediate energy domain.
In relativistic heavy-ion collisions, it was suggested to provide experimental evidence of $\alpha$ clustering in light nuclei in their ground state through collisions against heavy nuclei~\cite{broniowski2014signatures,bozek2014alpha}.
It can be expected that the shape of the created fireball in the transverse plane reflects the initial geometry of the light nucleus, because the interaction time is short enough to prevent the much slower nuclear excitation at ultrahigh collision energies.
In our previous works, we investigated initial geometry effects on elliptic and triangular flows and their fluctuations, Hanbury Brown-Twiss (HBT) correlations as well as electromagnetic fields in $^{12}$C + $^{197}$Au collisions~\cite{zhang2017nuclear,zhang2018collective,he2020clustering,MaLong,MaLong2,ChengYL,Dasgupta}.
It turns out that the difference of the second-order and third-order anisotropy could be clearly observed among spherical structure, clustered triangle and clustered chain.
However, the probes proposed so far are still limited.
If considering that the ground states of $\alpha$-conjugate nuclei may be composed of clustered configurations and non-clustered configurations, the previous observables are almost unable, or at least difficult, to provide the information on the mixture.

On the other hand, the machine learning (ML) revolution rose in the past decades.
In recent years, ML-based techniques have not only influenced many areas in the industry, but have also entered most scientific disciplines, including physics naturally~\cite{carrasquilla2017machine,tanaka2017detection,van2017learning,morningstar2017deep,CPL1,CPL2,BalNtC5,NST_ML}.
In heavy-ion collisions, impact parameter determination via neural networks dates back to the 1990s~\cite{bass1994neural,bass1996neural}.
Recently, ML has been used to identify the quantum chromodynamics (QCD) phase transitions~\cite{pang2018equation,Du,SteJHEP2019} and liquid-gas phase transitions~\cite{wang2020nuclear} as well as to determine the temperature in heavy-ion collisions~\cite{SongYD}.

Because the final-state anisotropy can reflect initial geometry and modern neural networks are powerful tools for extracting information from complex datasets, we expect that they could learn more features from anisotropic distributions, which may offer new ways to solve the current limitations.
In this work, we use a multi-phase transport (AMPT) model to simulate the central $^{12}$C/$^{16}$O + $^{197}$Au collisions at $\sqrt{S_{NN}} =$ 200 GeV.
After that, we train a Bayesian convolutional neural network (BCNN) using the 2D spectrum of azimuthal angle and transverse momentum to provide the information on the initial configuration of $^{12}$C/$^{16}$O.

The paper is organized as follows.
Sect.~\ref{SecAMPT} introduces the AMPT Model and algorithm for $^{12}$C/$^{16}$O initialization.
The Bayesian convolutional neural network we constructed is described in Sect.~\ref{SecBCNN}.
Results and discussion for training, validating the network and predicting on the test set with different proportions of initial configurations are then presented in Sect.~\ref{SecResults}.
Finally, a summary is given in Sect.~\ref{SecSummary}.

\section{\label{SecAMPT}The AMPT Model}
The AMPT model is a successful event generator for relativistic heavy-ion collisions, comprising four main stages: initialization, partonic interactions, hadronization and hadronic interactions~\cite{lin2005multiphase,Lin2021}.
There are two versions of the AMPT model, namely, the default version and the string melting version.
In both versions, the phase-space distributions of minijet partons and soft string excitations are initialized by the heavy ion jet interaction generator (HIJING) model~\cite{gyulassy1994hijing}.
Scatterings among partons are described by Zhang's parton cascade (ZPC)~\cite{zhang1998zpc}.
In the default version, minijet partons are recombined with the parent strings to form new excited strings after partonic interactions.
Then, these strings experience hadronization via the Lund string fragmentation model~\cite{sjostrand1994high}.
In the version of string melting, both excited strings and minijet partons are decomposed into partons.
And a naive quark coalescence model is responsible for the hadronization.
Finally, the interactions of the subsequent hadrons are modeled by a hadronic cascade, which is based on a relativistic transport (ART) model~\cite{li1995formation}.
The default AMPT model can give reasonable descriptions of rapidity and transverse momentum spectra in heavy-ion collisions.
However, it underestimates the elliptic flow observed at Relativistic Heavy Ion Collider (RHIC) because most of the energy produced in the overlap volume of heavy-ion collisions is in hadronic strings, thus not included in the parton cascade in the model~\cite{lin2002partonic}.
Hence, the string melting version is employed in this work.
AMPT has given good descriptions of different physics for relativistic heavy-ion collisions at both RHIC~\cite{lin2005multiphase} and LHC~\cite{AMPTGLM2016} energies, e.g., di-hadron azimuthal correlation~\cite{AMPTDiH,WangHai}, collective flow~\cite{STARFlowAMPT,AMPTFlowLHC}, strangeness production~\cite{SciChinaJinS}, chiral magnetic effect and so on~\cite{Zhao,Huang}.

\begin{figure}[htbp]
\centering
\includegraphics[width=0.45\textwidth]{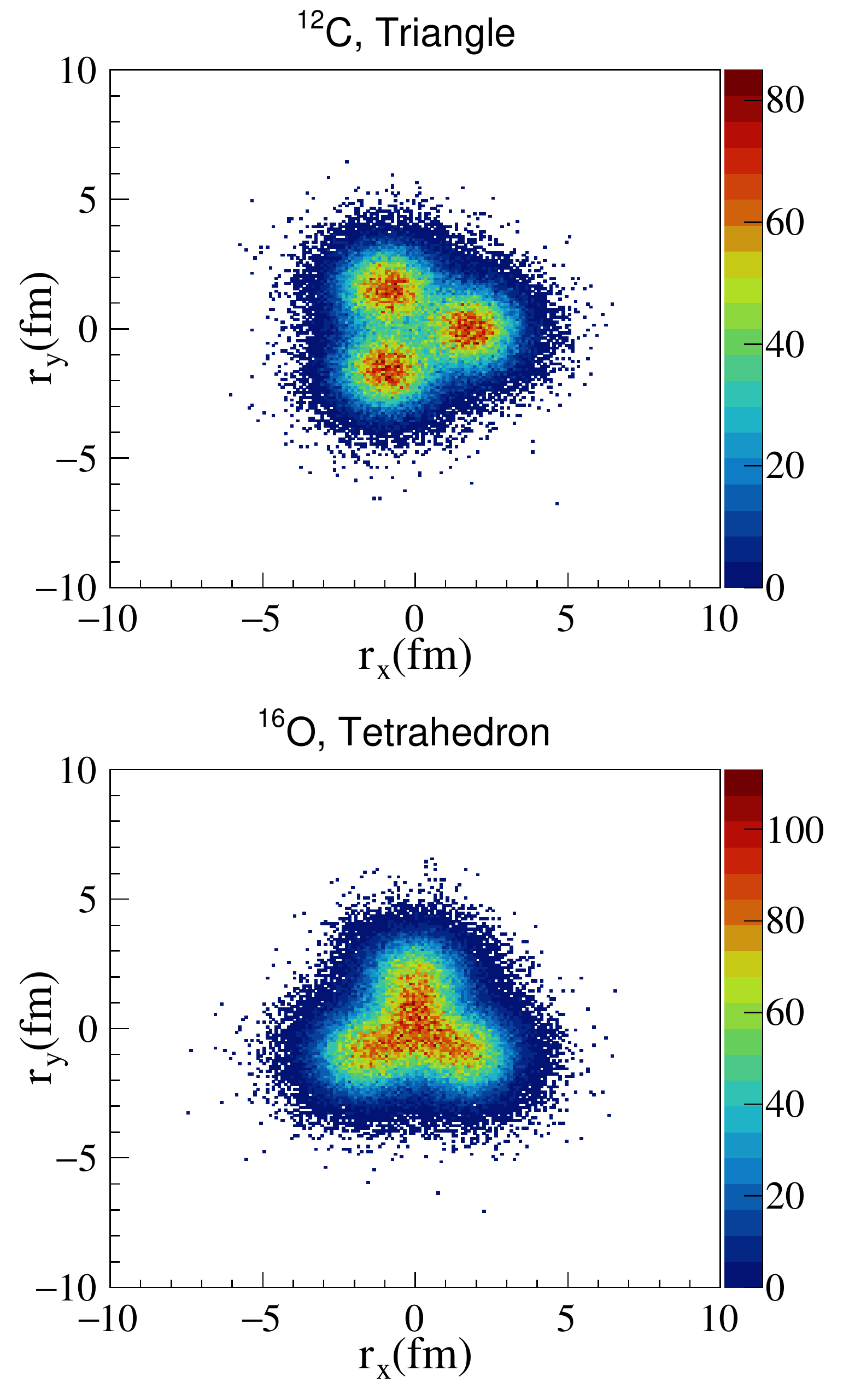}
\caption{Nucleon distributions in transverse plane for clustered $^{12}$C and $^{16}$O from 10000 events.}
\label{fig:configuration}
\end{figure}

In spite of the importance of clustering, models for relativistic heavy-ion collisions usually still initialize light nuclei by Woods-Saxon (W-S) distribution.
$^{12}$C is one of the typical examples where cluster and shell features coexist.
Both theoretical calculations and experimental evidence support a compact triangular structure for the ground state of $^{12}$C.
Therefore, we initialize $^{12}$C as three-$\alpha$ triangular structure besides the default Woods-Saxon distribution.
The centers of three $\alpha$ clusters are simply placed to form an equilateral triangle.
And the nucleons inside each $\alpha$ cluster are sampled by the three-parameter Woods-Saxon distribution in HIJING.
The side length is set to be 3.103 fm in order to match the measured value of the rms nuclear charge radius of $^{12}$C, which is 2.47 fm~\cite{angeli2013table}.
$^{16}$O with Woods-Saxon distribution and four-$\alpha$ tetrahedral configuration is also considered in this work.
Correspondingly, the edge length is set to be 3.416 fm.
After sampling 10000 events, Fig.~\ref{fig:configuration} visualizes the initial distributions of nucleons in transverse plane for clustered $^{12}$C and $^{16}$O.

\section{\label{SecBCNN}Bayesian Convolutional Neural Network}
There are basically three types of learning paradigms: supervised learning, unsupervised learning, and reinforcement learning.
Among the supervised learning methods, multilayer feed-forward neural networks, or more specifically multilayer perceptrons, stand behind the machine learning revolution of the past decade.
Inspired by the process of image processing in mammal’s visual cortex, convolutional neural networks (CNN) are important and powerful variants of neural networks.
Together with the development of computation and the accumulation of data, CNNs have greatly driven the progress of computer vision since 2012~\cite{krizhevsky2012imagenet}, although many key breakthroughs of algorithms can be traced back to the 80s and 90s last century~\cite{schmidhuber2015deep,lecun2015deep}.

Except an input and an output layer, the hidden layers of a CNN typically consist of convolutional layers, pooling layers and fully connected (FC) layers.
Convolutional layers are employed to extract useful features.
With the concepts of local receptive fields and shared weights, CNNs have much smaller number of parameters compared to multilayer perceptrons, and also translation invariance characteristics.
Batch Normalization (BN), a technique to make the training process faster and more stable by normalizing the input of each layer, is usually applied before the non-linearity~\cite{ioffe2015batch}.
The rectified linear unit (ReLU) is the most popular activation function for deep neural networks.
The purpose of pooling layers is to reduce the dimensions of the data.
Commonly, pooling takes the maximum or average value from each of a cluster of neurons at the prior layer.
Finally, fully connected layers are responsible for making the final prediction.
In brief, the whole network maps the extracted features to the desired output.

Regular neural networks only provide point estimates, while Bayesian neural networks utilize variational inference to learn the posterior distribution of the weights given the dataset, allowing them to measure uncertainty.
That is to find the parameters $\theta$ that minimize the Kullback-Leibler (KL) divergence of the variational posterior on the weights $q(\mathbf{w} \mid \theta)$ from the true Bayesian posterior $P(\mathbf{w} \mid \mathcal{D})$ on the weights~\cite{hinton1993keeping,graves2011practical}.
Therefore, the cost function is written as
\begin{equation}
\begin{aligned}
J(\mathcal{D}, \theta) =& \mathrm{KL}[q(\mathbf{w} \mid \theta) \| P(\mathbf{w})] \\
&- \mathbb{E}_{q(\mathbf{w} \mid \theta)}[\log P(\mathcal{D} \mid \mathbf{w})],
\label{CostFunction}
\end{aligned}
\end{equation}
where the first term measures the distance between the prior and variational posterior, and the negative log-likelihood measures the goodness of fit of the model.
Subsequently, the cost function can be optimized by an algorithm called ``Bayes by Backprop", which makes use of Monte Carlo sampling to obtain the unbiased estimates of the gradients of the cost function~\cite{blundell2015weight}.

\begin{table}[htbp]
\caption{\label{tab:BCNN}Architecture of Bayesian Convolutional Neural Network.}
\begin{tabular}{|l|l|l|}
    \hline
    No. & Layer             & Parameters\\ \hline
    0   & Input             & $1 \times 64 \times 64$ tensor\\
    1   & Conv2D            & 6 kernels ($3 \times 3$), BN, ReLU\\
    2   & MaxPool2D         & $2 \times 2$ kernel, stride 2\\
    3   & Conv2D            & 16 kernels ($3 \times 3$), BN, ReLU\\
    4   & MaxPool2D         & $2 \times 2$ kernel, stride 2\\
    5   & Conv2D            & 32 kernels ($3 \times 3$), BN, ReLU\\
    6   & MaxPool2D         & $2 \times 2$ kernel, stride 2\\
    7   & Conv2D            & 64 kernels ($3 \times 3$), BN, ReLU\\
    8   & MaxPool2D         & $2 \times 2$ kernel, stride 2\\
    9   & Flatten           & 1024 neurons\\
    10  & BayesianFC        & 256 neurons, BN, ReLU\\
    11  & BayesianFC        & 64 neurons, BN, ReLU\\
    12  & BayesianFC+Softmax& 2 outputs\\ \hline
\end{tabular}
\end{table}

The BCNN architecture employed in this work is listed in Table~\ref{tab:BCNN}.
The network consists of four convolutional layers, together with four pooling layers, and three subsequent Bayesian fully connected layers.
In total, there are about 0.58 million parameters.
Each convolutional layer applies ``same" padding and is followed by the BN layer, the ReLU function, and max pooling.
All convolutional filters are of size $3 \times 3$.
And it is a common choice for max pooling shape to be $2 \times 2$.
Filters scan through the outputs from previous layers and create feature maps as inputs for next layers.
The feature map from the last pooling layer is flattened.
After the features pass through the Bayesian fully connected layers, Softmax function will take logits and produce probabilities.
For binary classification, the prediction is made simply according to the Softmax output with threshold equal to 0.5.
Also, cross-entropy loss function, which corresponds to the second term in Eq.~(\ref{CostFunction}), is applied to measure the dissimilarity between predicted probability distributions and actual labels.
This network is realized in PyTorch framework~\cite{paszke2017automatic} with BLiTZ package~\cite{esposito2020blitzbdl}.

\section{\label{SecResults}Results and Discussion}
\subsection{Data Preparation and Training}
\begin{figure}[htbp]
\centering
\includegraphics[width=0.45\textwidth]{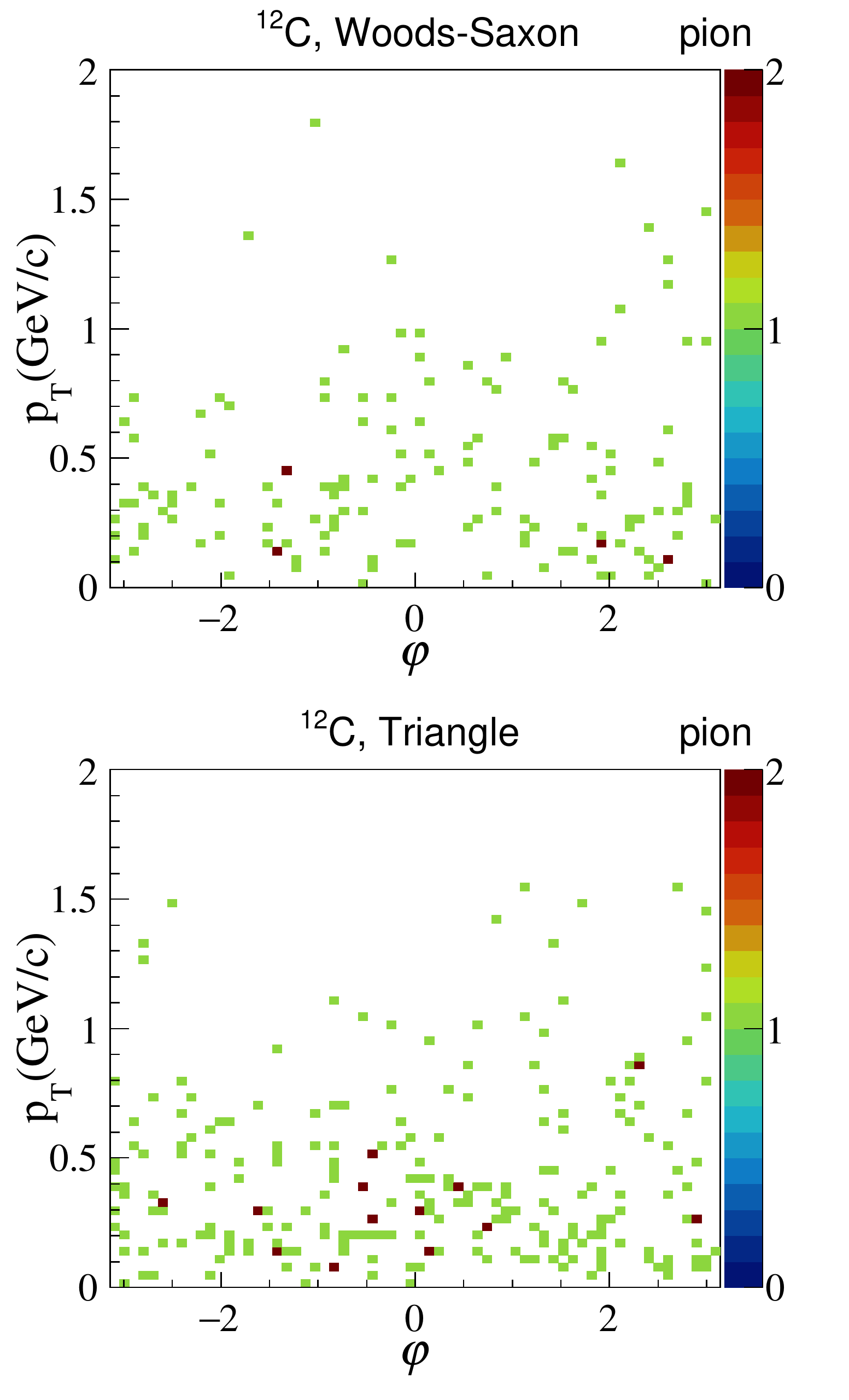}
\caption{Two-dimensional azimuthal angle and transverse momentum distributions of charged pions for non-clustered and clustered $^{12}$C from a single AMPT-generated $^{12}$C + $^{197}$Au collision event at $\sqrt{S_{NN}} =$ 200 GeV.}
\label{fig:spectrum1}
\end{figure}

With the AMPT model, $8 \times 10^5$ central $^{12}$C/$^{16}$O + $^{197}$Au collisions are generated for each configuration at $\sqrt{S_{NN}} =$ 200 GeV.
In a first step, the network was trained on event-by-event basis.
The network will certainly benefit from the whole final-state momentum space.
However, each event has different particle yields.
In order to keep each sample consistent and take advantage of the convolution neural network, 2-dimensional histograms of azimuthal angle and transverse momentum are designed as inputs.
Considering detection efficiency in experiments, charged pions with rapidity from -1 to 1 and transverse momentum from 0 to 2 GeV/$c$ are chosen because they are the most produced final-state particles in relativistic heavy-ion collisions.
For each event, a 2-dimensional histogram is filled with the selected pions.
Fig.~\ref{fig:spectrum1} displays typical pion distributions coming from one collision event generated with both non-clustered and clustered $^{12}$C.
Obviously, it is impossible for human being to distinguish the two cases.
The whole dataset consists of $1.6 \times 10^6$ histograms with $64 \times 64$ bins, labeled ``0" for the Woods-Saxon case and ``1" for the triangle/tetrahedron case.
For data preprocessing, each sample is standardized by the mean and standard deviation of its own bins.
This kind of sample-wise normalization could avoid the effect of multiplicity and accelerate the training process.
And the dataset is split into two parts, 80 percent for training and 20 percent for validation.
The network is trained by stochastic gradient descent with learning rate of 0.0003, momentum of 0.9 and minibatches of size 512 during 500 epochs.
To evaluate the performance of the classifier on the validation set, one could calculate the $F_1$ score defined as the harmonic mean of the precision $p$ and recall $r$, namely $F_1 = \frac{2pr}{p+r}$.
For each class, the precision is the number of true positives divided by the number of all positive predictions and the recall is the number of true positives divided by the number of all actually positive samples.
Model parameters that yield the best average $F_1$ score, or specifically macro $F_1$ score, on the validation set are saved for model evaluation afterwards.
However, the network cannot differentiate the initial configurations of $^{12}$C from the final-state transverse momentum space on event-by-event basis because the best accuracy on the validation set is just above $50\%$, which means the event-by-event fluctuation prevails over the initial geometry effect in such small colliding systems.
Chances are that the participant nucleon distributions of specific events of two configurations could be extremely similar to each other due to the initial-state fluctuation.

\begin{figure}[htbp]
\centering
\includegraphics[width=0.45\textwidth]{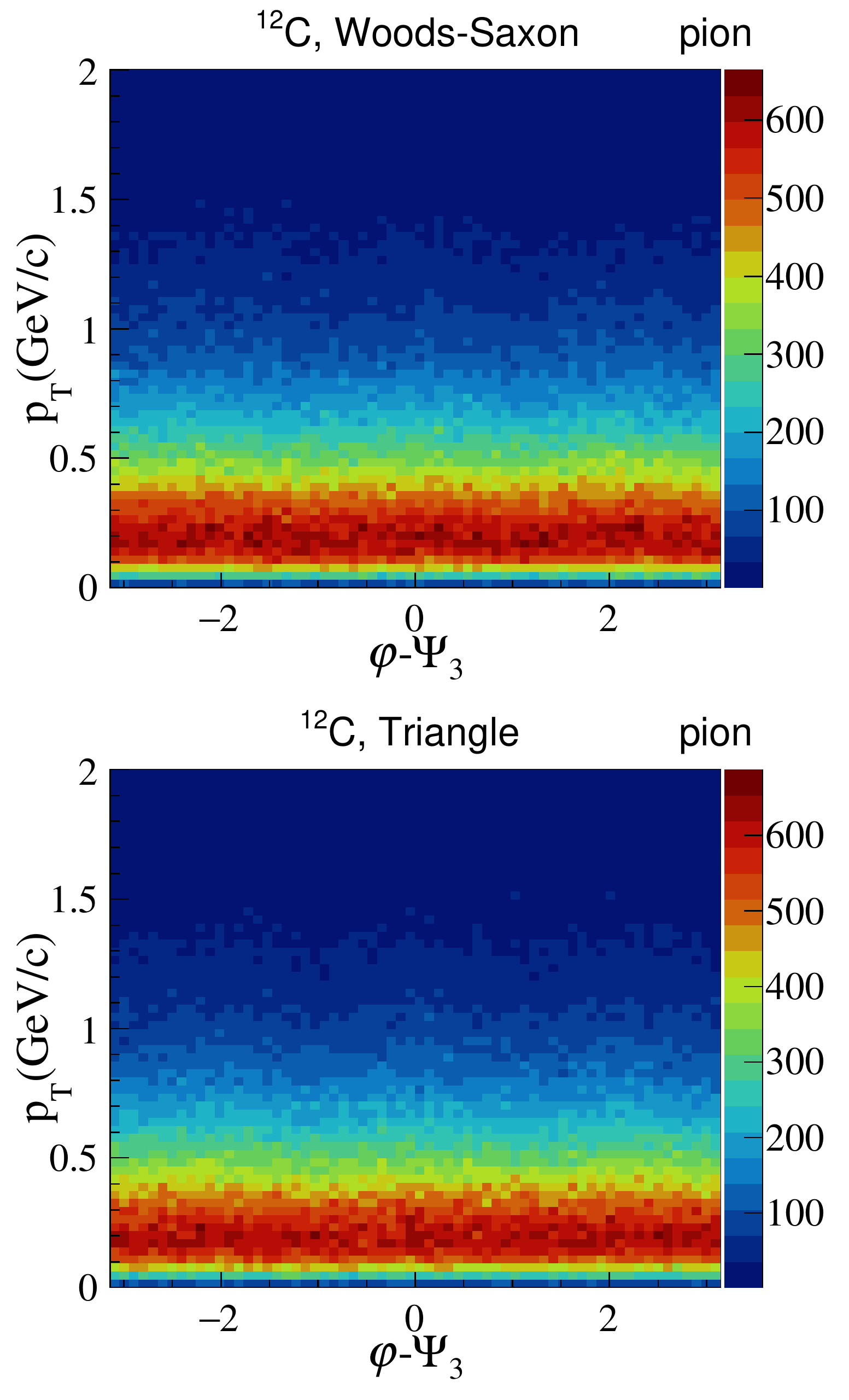}
\caption{Same as Fig.~\ref{fig:spectrum1} but for 4000 merged events which are rotated according to the event-by-event third-order participant plane.}
\label{fig:spectrum4000}
\end{figure}

\begin{figure}[htbp]
\centering
\includegraphics[width=0.45\textwidth]{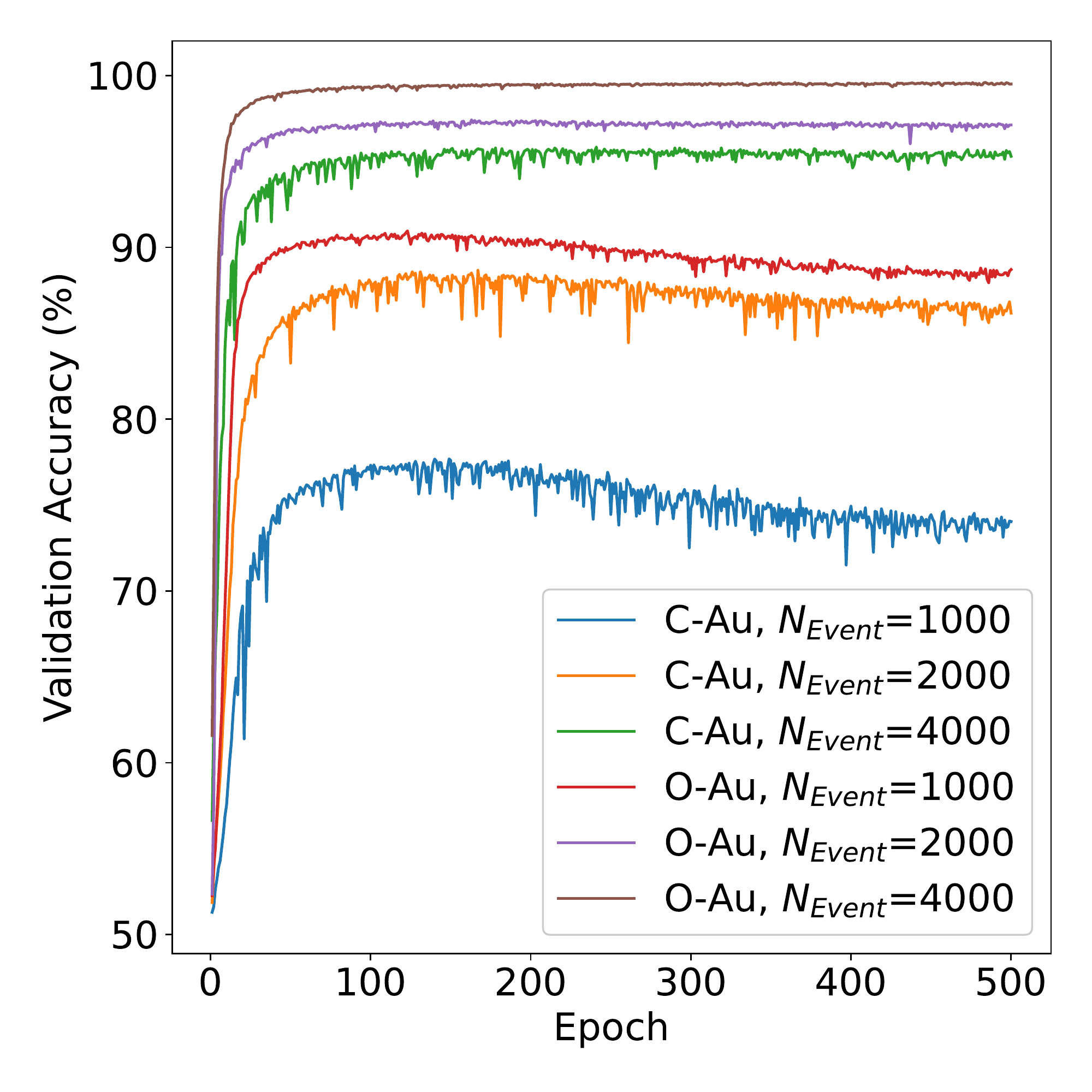}
\caption{Evolution of the validation accuracy during the training process for the dataset of two colliding systems $^{12}$C/$^{16}$O + $^{197}$Au with $N_{Event} =$ 1000, 2000 and 4000.}
\label{fig:training}
\end{figure}

Further, samples of multiple events were considered.
Events are randomly chosen to be merged as one sample after being rotated according to their third-order event planes respectively due to the triangular symmetry of $^{12}$C/$^{16}$O we configured.
The event plane estimation via Q vectors involves selection of particles and weights.
Therefore, for convenience, the event plane is estimated by the participant plane, which is defined by~\cite{voloshin2008elliptic,alver2010collision,lacey2011initial}
\begin{equation}
\Psi_n = \frac{\mathrm{atan2}\left(\frac{\left<r_{part}^2\sin\left(n\phi_{part}\right)\right>}{\left<r_{part}^2\cos\left(n\phi_{part}\right)\right>}\right)+\pi}{n},
\label{PartPlanDef}
\end{equation}
where $\Psi_n$ is the $n$th-order participant plane angle, $r_{part}$ and $\phi_{part}$ are radial coordinate and azimuthal angle of participants in the initial state, and $\left<\cdots\right>$ denotes the average over participants.
The typical spectra of 4000 merged events are depicted in Fig.~\ref{fig:spectrum4000}.
Even with the mergence, the samples of different configurations are still hardly discernible with the naked eye.
We denote the number of merged events as $N_{Event}$, and take $N_{Event}$ to be 1000, 2000 and 4000.
Together with two colliding systems, the performance of the network was investigated on these 6 datasets.
For each $N_{Event}$ and each configuration, $4 \times 10^4$ samples are generated from $8 \times 10^5$ events of raw data.
With the same training process mentioned above, the learning curves are summarized in Fig.~\ref{fig:training}.
Although the overall accuracies on the training set could easily reach $95\%$ or even $99\%$ for all 6 cases, those on the validation set have different performances.
The best three cases converge quite well, while there are peaks for the worst three cases.
The result is better for larger $N_{Event}$ because $N_{Event}$ reflects the competition between features and fluctuations.
If $N_{Event}$ is not large enough, the network will overfit the training set, and training on event-by-event basis is just a special case.
On the contrary, when event-by-event fluctuations are reduced, the network is capable of learning the features of the final state to predict the initial configuration.
For $^{12}$C with $N_{Event} =$ 4000 and $^{16}$O with $N_{Event} =$ 2000, validation accuracy reaches $95\%$ and $97\%$, respectively, and it is even $99\%$ for $^{16}$O with $N_{Event} =$ 4000.
With the same $N_{Event}$, the result of $^{16}$O has a better performance than that of $^{12}$C due to its 3D triangular symmetry.

\begin{figure*}[htbp]
\centering
\includegraphics[width=0.9\textwidth]{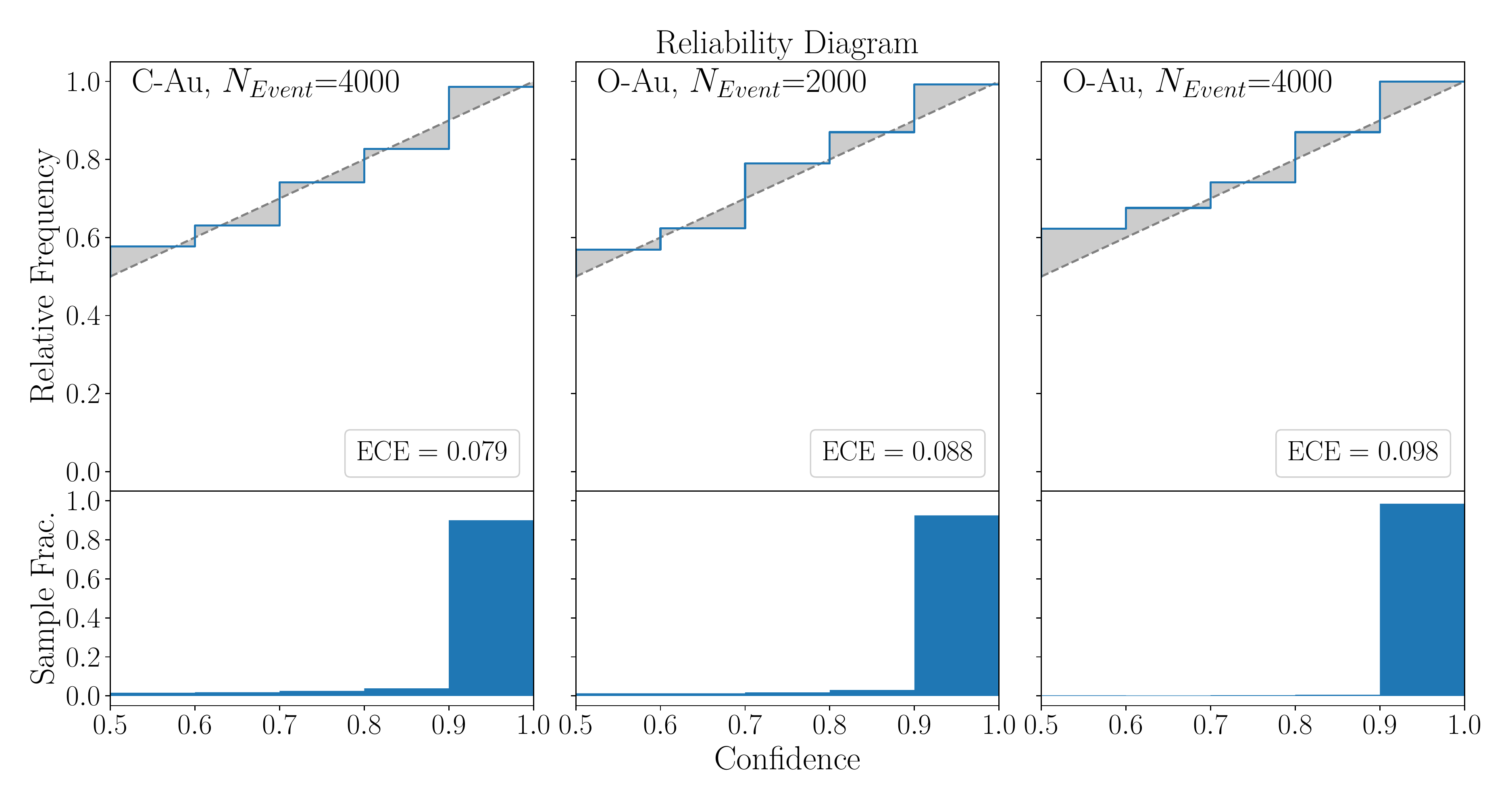}
\caption{The upper panels show the reliability diagrams for $^{12}$C with $N_{Event} =$ 4000, $^{16}$O with $N_{Event} =$ 2000 and $^{16}$O with $N_{Event} =$ 4000. The lower panels show the sample fractions in each confidence bin.}
\label{fig:ReliabilityDiagram}
\end{figure*}

Probabilistic classifiers output a probability distribution on target classes rather than just a label.
Naturally, calibrated confidence estimates are important for model interpretability.
A model is well-calibrated if the predicted probability reflects the ground truth correctness likelihood.
The reliability diagram, which plots the observed relative frequency as a function of confidence, is a common diagnostic for calibration.
Fig.~\ref{fig:ReliabilityDiagram} presents the reliability diagrams of the best three trained models as well as the sample fractions in each confidence bin.
The identity function (grey dashed line) means perfect calibration.
In order to quantify the miscalibration, one could define Expected Calibration Error (ECE) as the weighted average of the difference between observed relative frequency and confidence~\cite{naeini2015obtaining}.
Although there appears to be a negative correlation between ECE and validation accuracy, it turns out that all three models are already calibrated.
And in fact, several calibration methods were tried, including Platting scaling~\cite{platt1999probabilistic}, temperature scaling~\cite{guo2017calibration} and Gaussian process calibration~\cite{wenger2020non}, but none of them obviously improved these calibration curves.
This is reasonable because our network is quite shallow and for binary classification.
According to the bottom panels, predictions of most samples are within the last confidence bin, which is consistent with the high validation accuracy.

\subsection{Test on Mixed Dataset}
\begin{figure*}[htbp]
\centering
\includegraphics[width=0.9\textwidth]{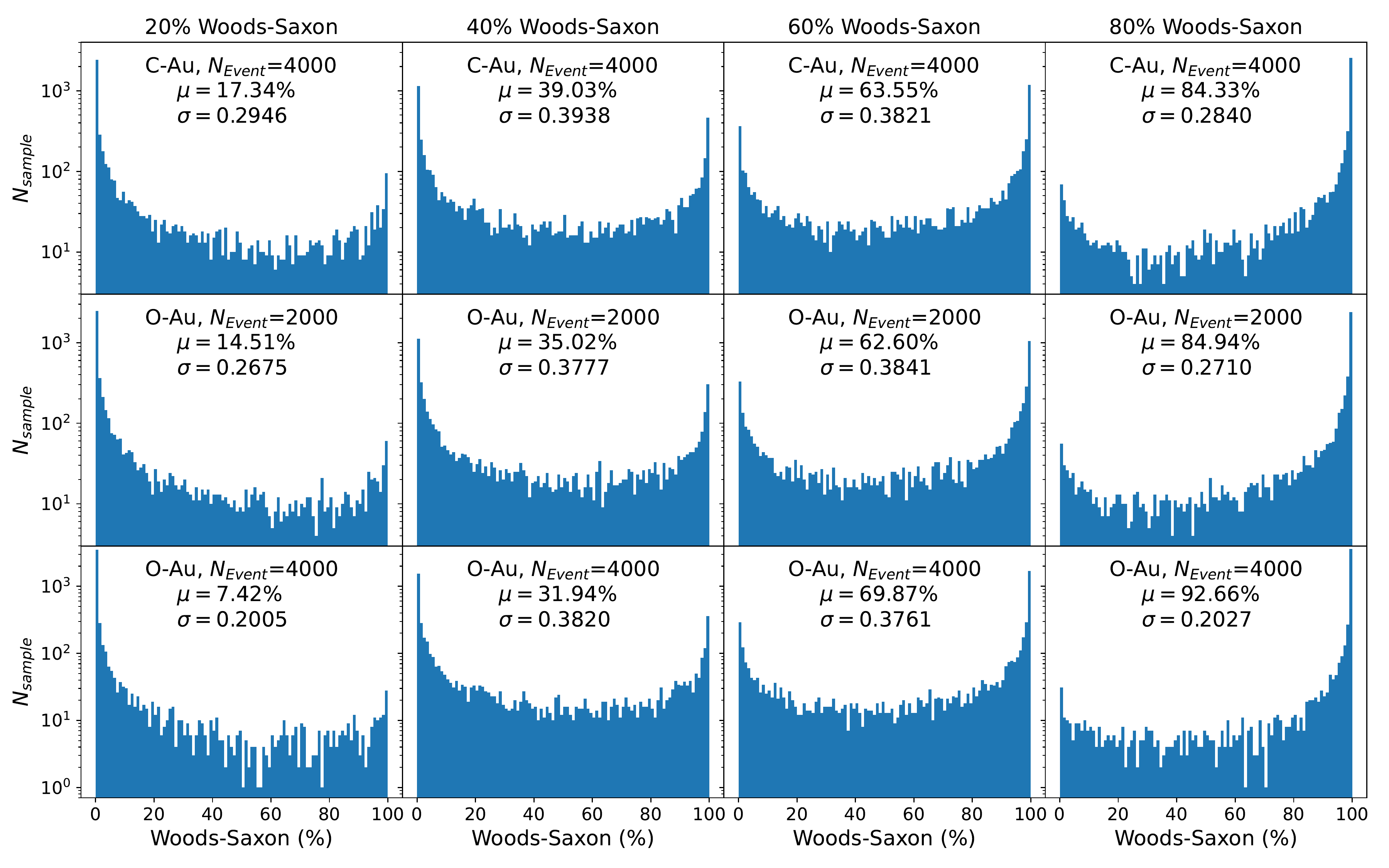}
\caption{Histograms of predictions on mixed samples for $^{12}$C with $N_{Event} =$ 4000 (upper panels), $^{16}$O with $N_{Event} =$ 2000 (middle panels) and $^{16}$O with $N_{Event} =$ 4000 (lower panels) with different mixing proportions of the Woods-Saxon configuration.}
\label{fig:mixtest}
\end{figure*}

\begin{figure*}[htbp]
\centering
\includegraphics[width=0.9\textwidth]{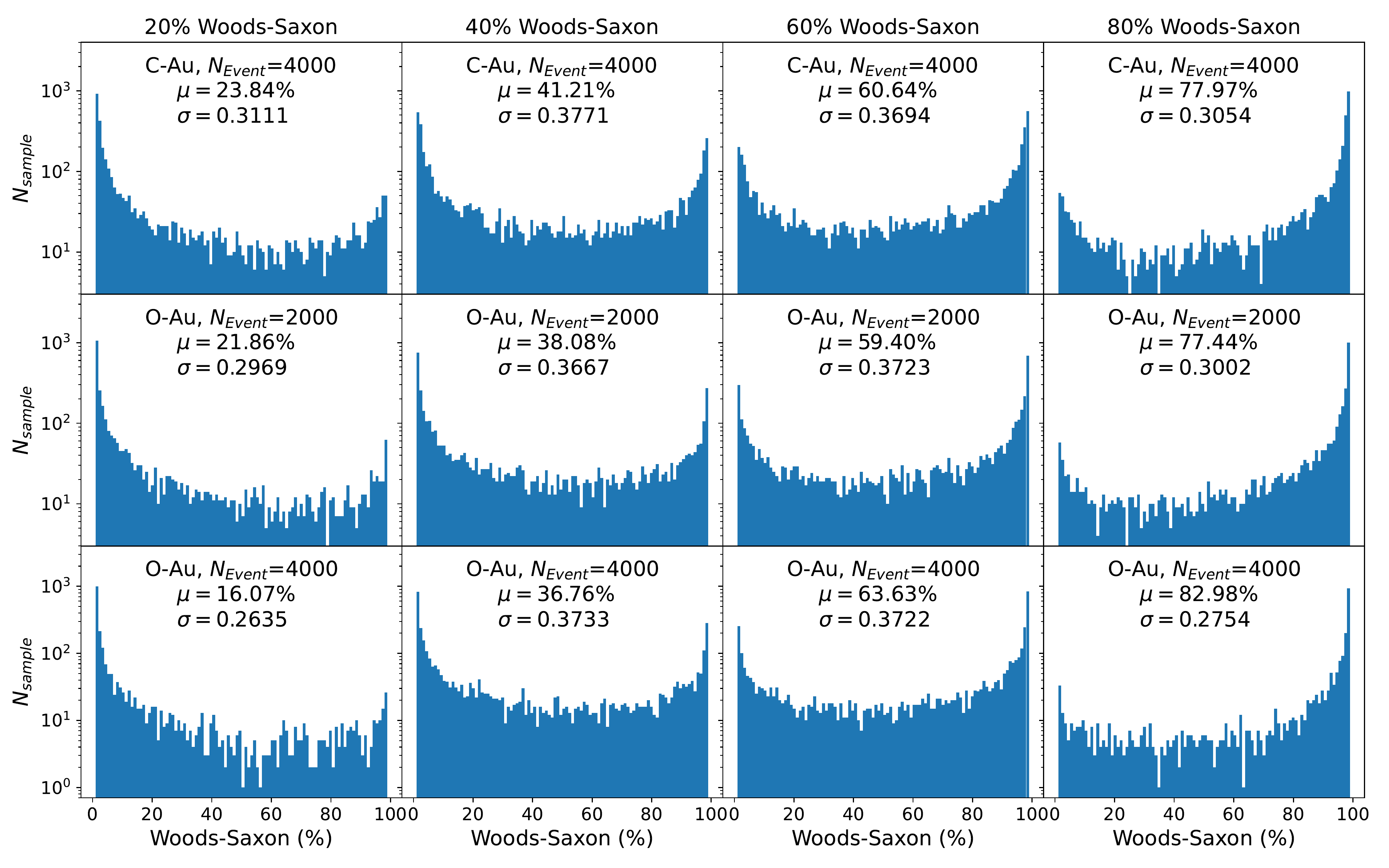}
\caption{Same as Fig.~\ref{fig:mixtest} but with confidence threshold of 0.99.}
\label{fig:mixtest_corrected}
\end{figure*}

\begin{figure*}[htbp]
\centering
\includegraphics[width=0.9\textwidth]{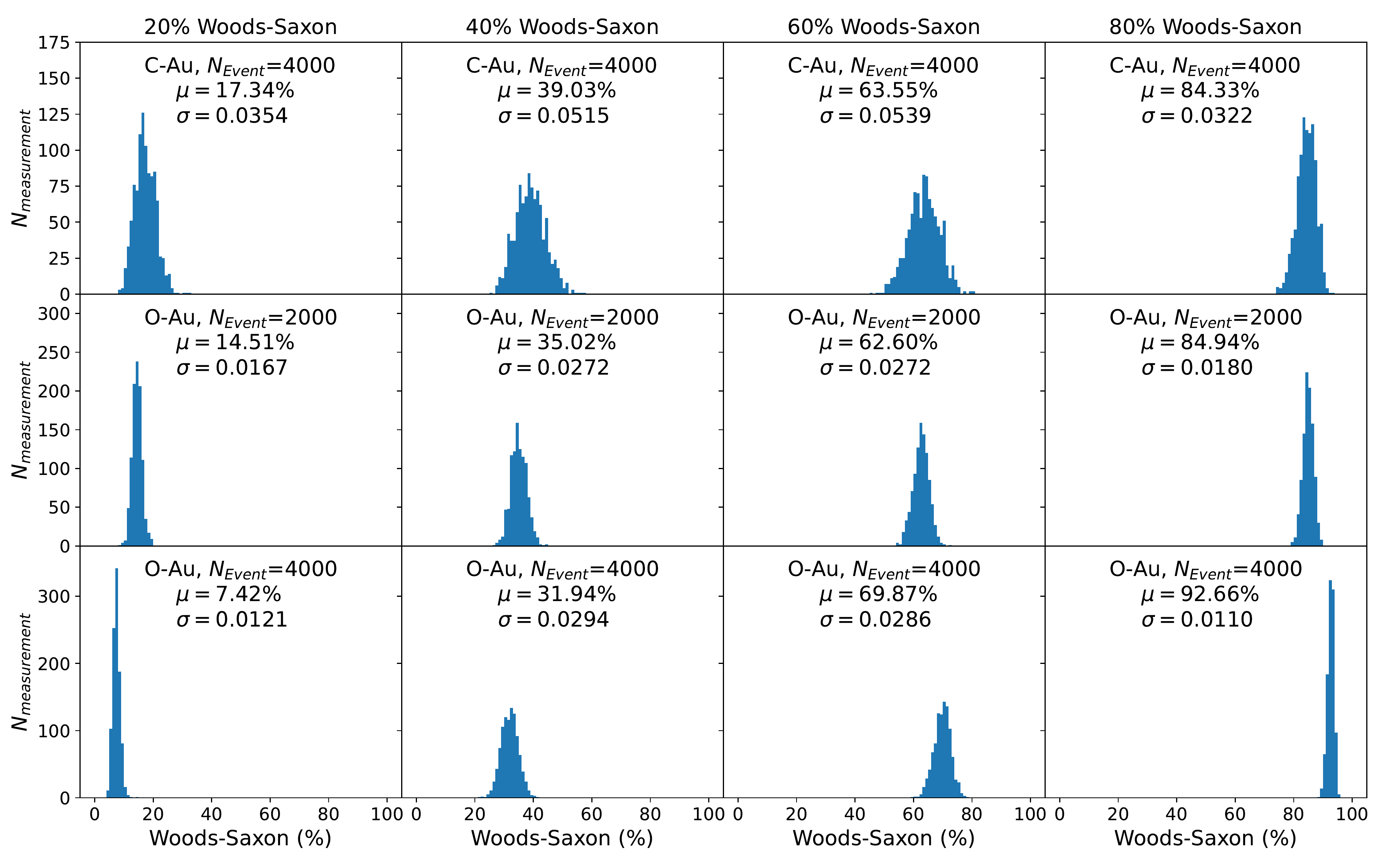}
\caption{Histograms of ``measurements" based on all mixed samples for $^{12}$C with $N_{Event} =$ 4000 (upper panels), $^{16}$O with $N_{Event} =$ 2000 (middle panels) and $^{16}$O with $N_{Event} =$ 4000 (lower panels) with different mixing proportions of the Woods-Saxon configuration.}
\label{fig:measure}
\end{figure*}

\begin{figure*}[htbp]
\centering
\includegraphics[width=0.9\textwidth]{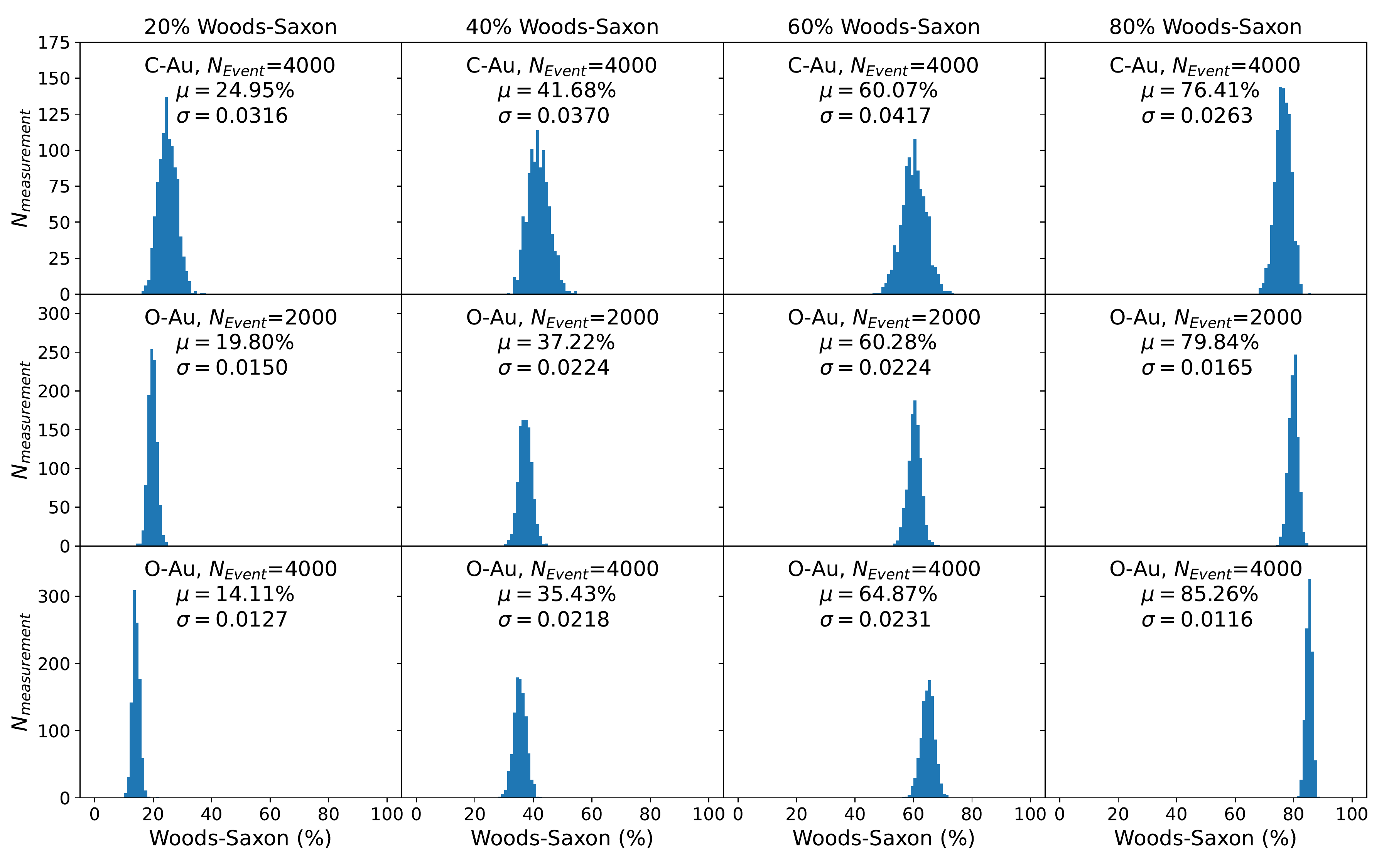}
\caption{Same as Fig.~\ref{fig:measure} but with confidence threshold of 0.999.}
\label{fig:measure_corrected}
\end{figure*}

Because the Softmax output could be interpreted as probability after calibration, it will be interesting to test our trained models on samples that are mixed with both configurations.
With the same $N_{Event}$, the proportion is selected as $20\%/80\%$, $40\%/60\%$, $60\%/40\%$ and $80\%/20\%$ for the Woods-Saxon and the triangle/tetrahedron cases, respectively.
From the best three trained classifiers, the predicted proportions of Woods-Saxon on 5000 mixed samples for each mixed dataset are shown in Fig.~\ref{fig:mixtest}.
As Bayesian neural networks sample the weights from the learnt posterior distributions every time before predicting, each mixed sample is predicted 1000 times and the mean value is taken to be filled in the histogram.
Even for the mixed samples of one single proportion, the average predictions form a broad distribution with the standard deviation $\sigma$.
And most predictions tend to be quite close to pure configurations.
Since the test set and the training set are not identically distributed, this broad distribution of predictions is acceptable.
The change in the distribution of data contained in training set and test set is called covariate shift, which is a huge and important topic in machine learning.
This kind of behavior means that sample by sample, the classifier would identify the features of the Woods-Saxon case or the $\alpha$-clustered case, reflecting the competition between initial geometry effects and fluctuations in mixed samples.
Furthermore, the means of the histograms are close to the true proportions, implying that the probability of the emergence of the features captured by the network is strongly related to the mixing proportion.
Thus, the average of the predictions on all mixed samples could be taken as the final prediction.
For $^{12}$C with $N_{Event} =$ 4000 and $^{16}$O with $N_{Event} =$ 2000, the deviations between true proportions and final predictions are almost within $5\%$, which is quite well considered the fluctuation and information loss in relativistic heavy-ion collisions.
Unfortunately, the deviations are around $10\%$ for $^{16}$O with $N_{Event} =$ 4000 in spite of the better accuracy on the validation set.
The almost perfect performance causes the overestimation on the mixed dataset.
As all three classifiers have the tendency to overestimate the proportion of the major component, it turns out that a simple confidence threshold could effectively correct the models.
For example, the confidence threshold is set to be 0.99, which means that if a predicted proportion of one single mixed sample is greater than 0.99 or less than 0.01, this prediction is ignored.
As shown in Fig.~\ref{fig:mixtest_corrected}, the confidence threshold of 0.99 considerably improves the ultimate overall results as the deviations are all within $4\%$ for three cases after correction.
There seems to be a larger correction when the initial imbalance is larger.
And it is worth mentioning that although the confidence threshold may result in differences between the weights of samples after averaging over Bayesian models, the subsequent average over samples is unweighted.
Considering the connection between the mixing proportion and overall predictions, if we think of the average value of the predictions on the whole mixed dataset as one ``measurement", in this case the 1000 predictions mentioned before are exactly 1000 ``measurements".
Fig.~\ref{fig:measure} presents the predicted proportions of Woods-Saxon of 1000 ``measurements" from three classifiers and for four mixed datasets, respectively, from which the means and the standard deviations of the models could be estimated.
When there is no cut, the means are the same as before.
Also, the corrected results are shown in Fig.~\ref{fig:measure_corrected}.
Note that the confidence threshold here is 0.999 because the weights are different from the previous situation.
Similarly, the means are closer to true values with all deviations almost no more than $5\%$ after correction.
The standard deviations are almost smaller as well.
Actually, since the three classifiers are independent of each other, the confidence thresholds are not necessarily the same values.

\begin{figure}[htbp]
\centering
\includegraphics[width=0.45\textwidth]{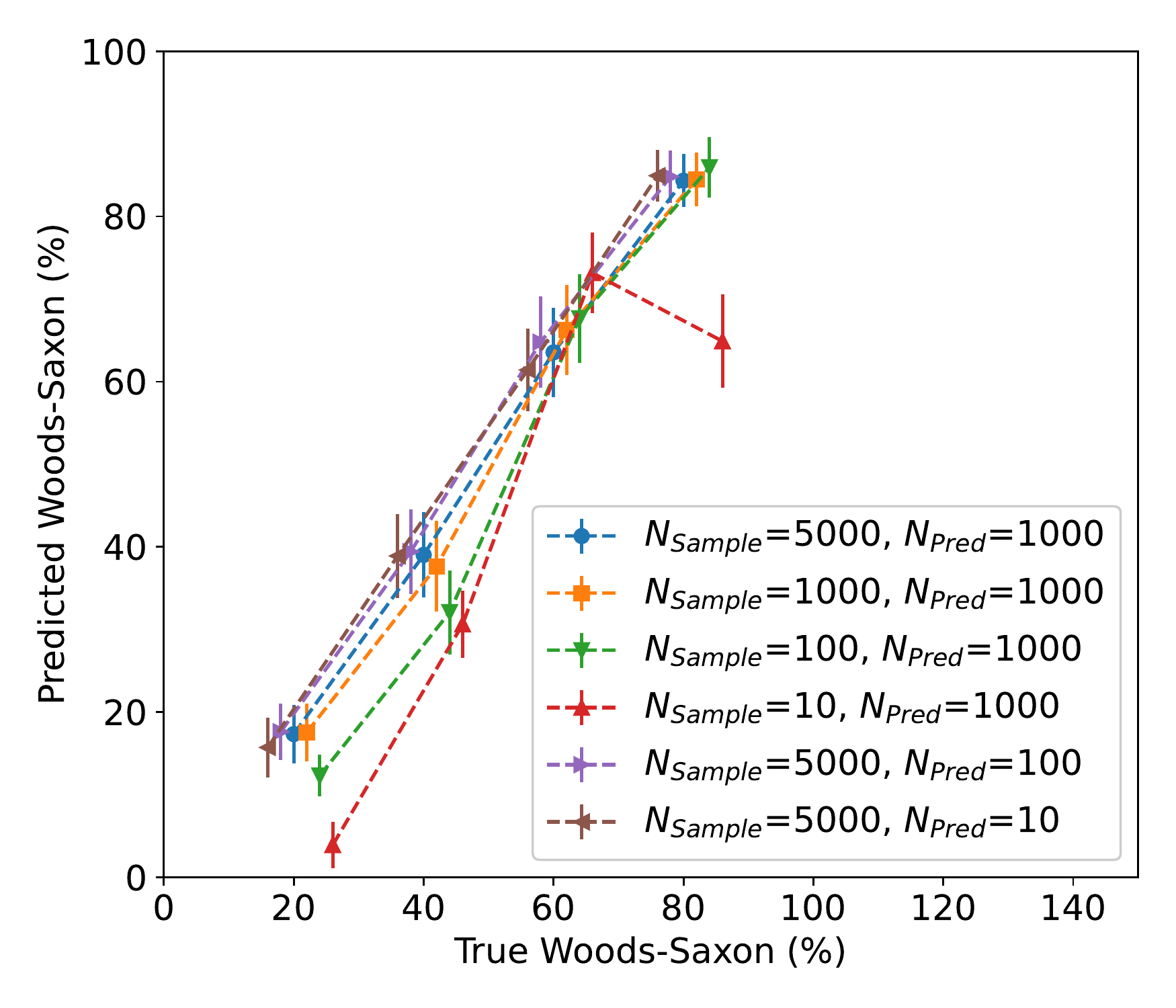}
\caption{Predicted proportion of the Woods-Saxon configuration versus true proportion of the Woods-Saxon configuration with different predicted times ($N_{Pred}$) and amounts of mixed samples ($N_{Sample}$). Here we take $^{12}$C with $N_{Event} =$ 4000 as an example. The deviation along the x-axis is for clarity.}
\label{fig:robust}
\end{figure}

Finally, the robustness of our predictions was investigated by taking $^{12}$C with $N_{Event} =$ 4000 as an example.
As illustrated in Fig.~\ref{fig:robust}, the predictions are almost stable if there are enough samples, which is 1000 in our case, and the times of ``measurement" hardly affect the predictions.
It can be seen that the standard deviation of models is reasonably small as well.

\subsection{Regression}
\begin{figure*}[htbp]
\centering
\includegraphics[width=0.9\textwidth]{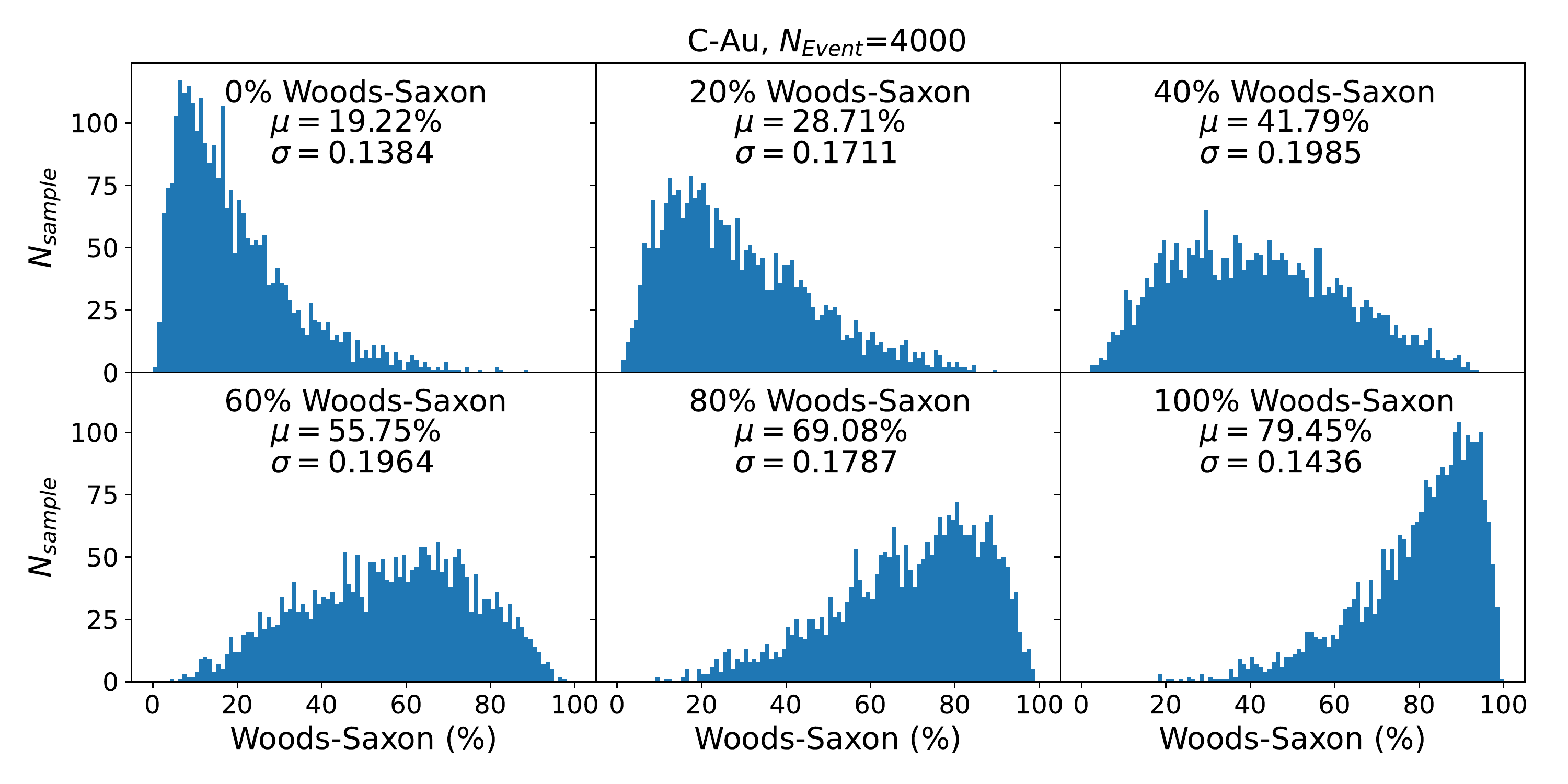}
\caption{Histograms of predictions on the validation set from a regression model for $^{12}$C with $N_{Event} =$ 4000.}
\label{fig:regression}
\end{figure*}

As a comparison, we also checked the performance of the regression model.
For the regression task, the architecture of BCNN is kept the same as before, except that the Softmax layer and cross-entropy loss are replaced by Sigmoid function and mean squared error, respectively.
The proportions of Woods-Saxon are taken as labels.
And there are 6 labels including pure and mixed configurations.
Similarly, the whole dataset has $8 \times 10^4$ samples and is split in the same way.
After training, the model with the least validation loss is evaluated.
Take $^{12}$C with $N_{Event} =$ 4000 as an example, Fig.~\ref{fig:regression} shows the predicted results on the validation set.
Although the peaks are related to the true proportions, the distributions of the predictions are still very broad.
Since the validation set and the training set are from the same original distribution this time, the performance of the regression model is not quite acceptable.
Additionally, the deviation between the mean value and the true proportion is larger when the initial imbalance is larger, because the Sigmoid function constrains the prediction to be from 0 to 1.
Therefore, the classification model achieved more satisfied performance than the regression one in this specific task.
The amount of data per label for regression is three times smaller than that for classification, which may partly explain the situation.

\subsection{Discussion and Prospect}
From this work, one can see that it is extremely difficult to extract signals of the initial state because fluctuations play such an important role in relativistic heavy-ion collisions.
Even after averaging over multiple events, the valuable information still depends on the mean of predictions.
However, notice that as these results were obtained from a simple BCNN and quite small datasets, this ML-based method seems to be fairly promising.
If models are trained multiple times, in different ways, or with different architectures, the average of multiple predictions may make the results more reliable.
For instance, the idea of residual network (ResNet)~\cite{he2016deep} could make the convolutional neural networks much deeper, and thus, state-of-the-art CNNs tend to use very small FC layers or even no FC layer, which significantly decreases the number of parameters.
As deep learning is data-hungry, more data and more sophisticated algorithms will certainly lead to greater achievements.
For example, smaller and denser grid with corresponding larger $N_{Event}$ will preserve more detailed information, hence allowing the network to learn more.
Heavier particles, such as kaons and protons, could offer better connections to the initial state, although larger $N_{Event}$ may be needed to reduce the fluctuations.
Actually, we trained our BCNN on 2D histograms of protons for $^{12}$C with $N_{Event} =$ 4000 as well.
However, the validation accuracy is only slightly over $70\%$.
While we tried a very small ResNet, which has only about 19 thousand parameters, it turns out that the training and validation results from protons are almost as good as those from pions.
Note that as the yield of proton is much lower than pion, it indicates that the samples of protons contain both stronger signals and fluctuations.
Moreover, carefully tuning the cuts of transverse momentum and rapidity may reveal which momentum space region is more sensitive to the initial state geometry and fluctuations.
On the other hand, the possibility of the method to be directly applied to experiments is obviously charming.
Although the event plane is estimated by the participant plane in this work, the results from the final-state-estimated event plane are expected to be consistent.
For experimental convenience and reliability, the most important goal of further studies is to find different physics observables as inputs to reduce fluctuations and improve the success of event-by-event analyses.
After being designed and trained properly, the model could make predictions on experimental data and provide the information on real nuclei, at least being a meaningful reference for theoretical physicists.
Experimental data are understood through theoretical models.
With the help of machine learning, experimental results could be easily connected to specific theoretical hypotheses.
Different classifiers can also be obtained by training on datasets from different models, e.g., UrQMD and hydrodynamics models, and their predictions on real data can be compared.
This process will improve our understanding of either structures of nuclei or dynamics of different models.
Now that the RHIC-STAR and LHC-ALICE experiments have been promoting collisions of small systems, such as $^{16}$O + $^{16}$O, the interface between ML and heavy-ion physics needs further investigation.

\section{\label{SecSummary}Summary}
In summary, we studied the initial clustering effect in relativistic heavy-ion collisions via a machine-learning-based method.
Configurations of $^{12}$C/$^{16}$O are initialized as $\alpha$-clustered triangle/tetrahedron and Woods-Saxon distribution of nucleons, and then the $^{12}$C/$^{16}$O + $^{197}$Au central collisions at $\sqrt{s_{NN}} =$ 200 GeV are simulated by the string melting AMPT model.
Based on 2D azimuthal-angle-transverse-momentum spectra of charged pions in the final state, a simple Bayesian convolutional neural network is trained.

From the results, we learn that the effect of event-by-event identification of initial $^{12}$C configurations is unsurprisingly unsatisfactory, because the initial fluctuation is so large in such small systems that events of different configurations could have similar participant nucleon distributions.
Despite the failure on event-by-event basis, the Bayesian convolutional neural network demonstrated excellent performance on classification of initial $^{12}$C/$^{16}$O configurations in the AMPT-generated heavy-ion collisions on multiple-event basis when the number of merged events $N_{Event}$ is large enough.
Especially, the trained model was tested on the mixed datasets with different proportions of the Woods-Saxon configuration, i.e., $20\%$, $40\%$, $60\%$ and $80\%$, and averagely, it succeeded in forecasting the combined proportions with the errors around $5\%$ with proper $N_{Event}$.
Further improvement was achieved after a simple confidence threshold was applied to the predictions.
On the other hand, the current results are still limited to the framework of averaging.
Based on greater computational power, further investigations could consider more complicated datasets and elaborately designed networks, which may make use of the whole final-state information, in order to deal with fluctuations.

More importantly, this framework can be easily extended to experimental data.
Machine learning could be a bridge between theoretical models and experimental results.
Due to the magnificent performance and experimental feasibility, observables combined with machine learning could bring more insights than traditional probes.
From the results, ML-based analyses exhibit the strong capability of learning features from fluctuations, which is helpful and useful to other topics in complex systems.

\begin{acknowledgements}
J. He thanks Dr. Rui Wang for useful discussions and Dr. Chen Zhong for maintenance of GPU farm.
This work was supported in part by  the Guangdong Major Project of Basic and Applied Basic Research No. 2020B0301030008, the National Natural Science Foundation of China under contract Nos. 11890710, 11890714, 11875066, and 11775288, and the National Key R\&D Program of China under Grant Nos. 2016YFE0100900 and 2018YFE0104600, the Strategic Priority Research Program of the CAS under Grants No. XDB34000000.
\end{acknowledgements}
\bibliography{MyBibliography}

\end{document}